\documentclass[aps,prb,twocolumn,superscriptaddress,longbibliography]{revtex4-2}

\usepackage{amsmath,amssymb,amsfonts}
\usepackage{graphicx}
\usepackage{bm}
\usepackage{hyperref}
\usepackage{color}
\usepackage{mathtools}

\newcommand{\bk}{\mathbf{k}}
\newcommand{\bg}{\mathbf{g}}
\newcommand{\BZ}{\mathrm{BZ}}

\begin{document}

\title{Symmetry selection rules for the intrinsic nonlinear thermal Hall effect in altermagnets: Role of quantum metric and $C_{2}$ rotational symmetry}

\author{Gunn Kim}
\affiliation{Department of Physics, Sejong University, Seoul 05006, South Korea}

\date{\today}

\begin{abstract}
We establish symmetry-based selection rules for the intrinsic nonlinear thermal Hall effect driven by the quantum metric in altermagnets. We show that a nonvanishing nonlinear thermal Hall conductivity $\kappa_{xyy}$ requires three conditions: (i) a nontrivial quantum metric, (ii) breaking of mirror symmetry $M_{x}$, and (iii) breaking of twofold rotational symmetry $C_{2}$. Using tight-binding models on a square lattice, we demonstrate that $d$-wave altermagnets naturally break $C_{2}$ through parity-mixing orbital hybridizations, while $g$-wave systems preserve $C_{2}$, forcing the response to vanish identically. Step-by-step Taylor expansions and explicit unitary matrix proofs establish these results. Our framework provides predictive power for material selection and lays the groundwork for nonlinear spin-caloritronic devices.
\end{abstract}

\maketitle

\section{Introduction}
Altermagnets represent a newly identified magnetic phase characterized by collinear antiferromagnetic order that supports strong, momentum-dependent spin splitting without macroscopic time-reversal symmetry breaking \cite{Smejkal2022a, Smejkal2022b, Smejkal2020, Mazin2022, Mazin2023, Hayami2019, Ahn2019}. Unlike conventional antiferromagnets where Kramers degeneracy is enforced globally, the unique spin splitting in altermagnets exhibits angular momentum structures classified by even harmonics such as $d$-wave ($l=2$), $g$-wave ($l=4$), and $i$-wave ($l=6$) \cite{Smejkal2022b, Smejkal2022c}.

The theoretical predictions of altermagnetism have sparked intense experimental efforts. These led to the direct observation of giant spin splittings via angle-resolved photoemission spectroscopy (ARPES) in various candidate materials including MnTe \cite{Krempasky2024, Lee2024, Osumi2024}, CrSb \cite{Reimers2024}, and Mn$_{5}$Si$_{3}$ \cite{Krempasky2024_Mn5Si3}. Consequently, the interplay between this unconventional magnetic order and the underlying crystal symmetry generates a rich variety of transport phenomena. Experimental and theoretical studies have reported the anomalous Hall effect \cite{Feng2022, Reichlova2024, Smejkal2020, Betancourt2023}, the spin-splitter effect \cite{Gonzalez2023, Shao2024, Bai2023}, and novel magnetoresistance behaviors \cite{Hajlaoui2024, Zhou2024, Chen2024}.

Beyond linear transport, nonlinear responses offer powerful probes of geometric and topological properties in quantum materials. In the electrical domain, the second-order nonlinear Hall effect has been extensively studied, primarily driven by two intrinsic geometric quantities. These are the Berry curvature dipole \cite{Sodemann2015, Ma2019, Kang2019, Du2021} and the Berry connection polarizability (BCP) \cite{Gao2014, Liu2021, Wang2023, Lau2021, Kaplan2024}. In systems with combined $\mathcal{PT}$ symmetry where the Berry curvature vanishes, the BCP provides the leading intrinsic contribution to the nonlinear Hall signal \cite{Liu2021, Wang2023, Gao2023}. The BCP is intrinsically linked to the quantum metric \cite{Provost1980, Resta2011}.

Recently, the discovery of a large magnetic nonlinear Hall effect (MNLHE) in the $d$-wave altermagnet Mn$_{5}$Si$_{3}$ \cite{Han2025} has demonstrated that nonlinear electrical transport serves as a sensitive experimental signature of altermagnetic symmetry breaking \cite{Wang2024_NHE, Zhang2024_NHE, Gao2024_NHE}. Despite rapid progress in the electrical domain, the extension of intrinsic nonlinear Hall physics to the thermal domain remains unexplored. While researchers have theoretically investigated the linear thermal Hall effect in altermagnets \cite{Bose2022, Zhou2024_Thermal, Mokrousov2024}, the geometrical origin of the second-order nonlinear thermal Hall effect requires systematic clarification. The foundation for such geometric thermal transport relies on Luttinger's formal prescription \cite{Luttinger1964, Qin2011, Xiao2006, Yu2019, Zeng2022}. However, the explicit symmetry constraints for altermagnets have not been fully mapped out.

In this work, we provide a step-by-step framework that establishes the central symmetry selection rules for the intrinsic nonlinear thermal Hall effect in altermagnets. By detailing every derivation, we prove that the twofold rotational symmetry ($C_{2}$) acts as the gatekeeper for macroscopic geometric thermal transport. This explains why materials like Mn$_{5}$Si$_{3}$ succeed where highly symmetric systems fail.

\section{Model Formulation and Quantum Metric Derivation}

\subsection{Algebraic derivation of the quantum metric}
We consider a time-reversal-symmetric two-band effective Hamiltonian:
\begin{equation}
H(\bk) = \bg(\bk)\cdot\bm{\sigma} = g_{x}(\bk)\sigma_{x} + g_{z}(\bk)\sigma_{z},
\end{equation}
where we set $g_{y}=0$. The band energies are $\epsilon_{\pm}(\bk) = \pm G(\bk)$ with $G = \sqrt{g_{x}^{2} + g_{z}^{2}}$.

The quantum metric $g_{ab}$ captures the geometric distance between neighboring Bloch states. We start from the general expression \cite{Provost1980}:
\begin{equation}
g_{ab} = \frac{1}{4G^{4}} [ G^{2} (\partial_{a} \bg \cdot \partial_{b} \bg) - (\bg \cdot \partial_{a} \bg)(\bg \cdot \partial_{b} \bg) ].
\end{equation}
We explicitly evaluate the inner products for our two-component vector $\bg = (g_{x}, 0, g_{z})$. First, the scalar product of derivatives is $\partial_{a} \bg \cdot \partial_{b} \bg = \partial_{a} g_{x} \partial_{b} g_{x} + \partial_{a} g_{z} \partial_{b} g_{z}$. Second, the product between the vector and its derivative is $\bg \cdot \partial_{a} \bg = g_{x} \partial_{a} g_{x} + g_{z} \partial_{a} g_{z}$. Substituting these into the numerator and expanding the energy gap term $G^{2} = g_{x}^{2} + g_{z}^{2}$, we obtain:
\begin{align}
\text{Num} &= (g_{x}^{2} + g_{z}^{2})(\partial_{a} g_{x} \partial_{b} g_{x} + \partial_{a} g_{z} \partial_{b} g_{z}) \nonumber \\
&- (g_{x} \partial_{a} g_{x} + g_{z} \partial_{a} g_{z})(g_{x} \partial_{b} g_{x} + g_{z} \partial_{b} g_{z}).
\end{align}
We expand the brackets to identify the canceling terms. The expanded numerator is:
\begin{align}
\text{Num} &= g_{x}^{2} \partial_{a} g_{x} \partial_{b} g_{x} + g_{x}^{2} \partial_{a} g_{z} \partial_{b} g_{z} + g_{z}^{2} \partial_{a} g_{x} \partial_{b} g_{x} \nonumber \\
&+ g_{z}^{2} \partial_{a} g_{z} \partial_{b} g_{z} 
- g_{x}^{2} \partial_{a} g_{x} \partial_{b} g_{x} - g_{x} g_{z} \partial_{a} g_{x} \partial_{b} g_{z} \nonumber \\
&- g_{z} g_{x} \partial_{a} g_{z} \partial_{b} g_{x} - g_{z}^{2} \partial_{a} g_{z} \partial_{b} g_{z}.
\end{align}
The terms $g_{x}^{2} \partial_{a} g_{x} \partial_{b} g_{x}$ and $g_{z}^{2} \partial_{a} g_{z} \partial_{b} g_{z}$ cancel exactly. The surviving terms are:
\begin{align}
\text{Num} &= g_{x}^{2} \partial_{a} g_{z} \partial_{b} g_{z} + g_{z}^{2} \partial_{a} g_{x} \partial_{b} g_{x} \nonumber \\
&- g_{x} g_{z} (\partial_{a} g_{x} \partial_{b} g_{z} + \partial_{a} g_{z} \partial_{b} g_{x}).
\end{align}
We now verify that this equals the factored form $(g_{z} \partial_{a} g_{x} - g_{x} \partial_{a} g_{z})(g_{z} \partial_{b} g_{x} - g_{x} \partial_{b} g_{z})$. Expanding the product gives $g_{z}^{2} \partial_{a} g_{x} \partial_{b} g_{x} - g_{z} g_{x} \partial_{a} g_{x} \partial_{b} g_{z} - g_{x} g_{z} \partial_{a} g_{z} \partial_{b} g_{x} + g_{x}^{2} \partial_{a} g_{z} \partial_{b} g_{z}$, which matches term by term. Therefore, the quantum metric takes the compact form:
\begin{equation}
g_{ab} = \frac{(g_{z} \partial_{a} g_{x} - g_{x} \partial_{a} g_{z})(g_{z} \partial_{b} g_{x} - g_{x} \partial_{b} g_{z})}{4(g_{x}^{2} + g_{z}^{2})^{2}}.
\label{eq:qm_derived}
\end{equation}
This derivation shows that a non-zero $g_{ab}$ strictly requires interband coupling driven by orbital mixing or spin-orbit coupling.

\subsection{$d$-wave altermagnet}
Motivated by the orthorhombic phase of Mn$_{5}$Si$_{3}$ \cite{Reichlova2024}, we construct the $d$-wave exchange splitting on a square lattice:
\begin{equation}
g_{z}(\bk) = \Delta(\cos k_{x} - \cos k_{y}).
\label{eq:gz_dwave}
\end{equation}
To generate a nontrivial quantum metric, we introduce an interband coupling consisting of a parity-even orbital mixing term and a parity-odd spin-orbit coupling term:
\begin{equation}
g_{x}(\bk) = 2t_{1}\sin k_{x} \sin k_{y} + \lambda \sin k_{y}.
\label{eq:gx_dwave}
\end{equation}

\subsection{True 2D $g$-wave altermagnet and the $O(k^{4})$ proof}
To properly contrast the $d$-wave case, we construct a mathematically rigorous 2D $g$-wave model on the same square lattice:
\begin{equation}
g_{z}(\bk) = \Delta \sin k_{x} \sin k_{y} (\cos k_{x} - \cos k_{y}).
\label{eq:gz_gwave}
\end{equation}
We define the interband coupling as $g_{x}(\bk) = \lambda \sin k_{y}$. We prove the $g$-wave nature by performing a Taylor expansion around the $\Gamma$ point. The sine terms expand as $\sin k_{x} \approx k_{x}$ and $\sin k_{y} \approx k_{y}$. The cosine difference expands as follows:
\begin{equation}
\cos k_{x} - \cos k_{y} \approx (1 - k_{x}^{2}/2) - (1 - k_{y}^{2}/2) = -\frac{1}{2}(k_{x}^{2} - k_{y}^{2}).
\end{equation}
We transform the momentum to polar coordinates where $k_{x} = k \cos \theta$ and $k_{y} = k \sin \theta$. We evaluate the product:
\begin{align}
g_{z}(\bk) &\approx \Delta (k \cos \theta) (k \sin \theta) \left[ -\frac{1}{2}(k^{2} \cos^{2} \theta - k^{2} \sin^{2} \theta) \right] \nonumber \\
&= -\frac{1}{2}\Delta k^{4} (\cos \theta \sin \theta) (\cos 2\theta).
\end{align}
We apply the trigonometric identity $\sin 2\theta = 2 \sin \theta \cos \theta$:
\begin{equation}
g_{z}(\bk) \approx -\frac{1}{4}\Delta k^{4} \sin 2\theta \cos 2\theta.
\end{equation}
We apply the identity again to obtain the final result:
\begin{equation}
g_{z}(\bk) \approx -\frac{1}{8}\Delta k^{4} \sin 4\theta.
\label{eq:gwave_taylor}
\end{equation}
This step-by-step expansion guarantees that the leading term is exactly $O(k^{4})$ with a fourfold angular dependence. Crucially, $g_{z}(-\bk) = g_{z}(\bk)$, which makes it a pure even function.

\begin{figure}[t]
\centering
\includegraphics[width=\columnwidth]{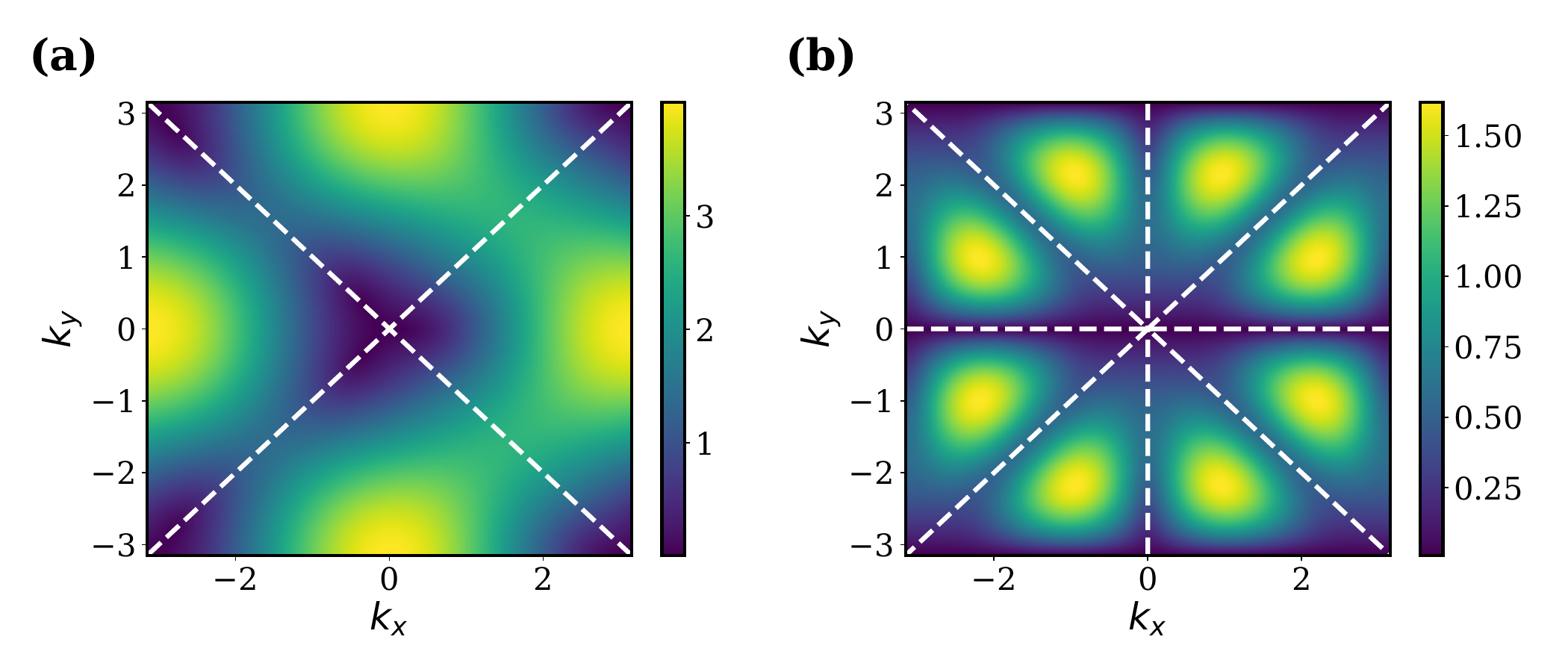}
\caption{Comparison of the energy gap $2G(\bk)$ and nodal lines (white dashed lines). (a) The $d$-wave model forms nodal lines along $k_{x} = \pm k_{y}$. (b) The $g$-wave model possesses a complex eightfold nodal structure. Parameters: $\Delta = 1.0$, $t_{1} = 0.5$, $\lambda = 0.3$.}
\label{fig:nodal_lines}
\end{figure}

\section{Symmetry Selection Rules and Matrix Proofs}

\subsection{Mirror symmetry breaking}
Under a mirror reflection $M_{x}$ mapping $x$ to $-x$, each spatial index $x$ in the rank-3 tensor $\kappa_{abc}$ acquires a factor of $(-1)$, so that $\kappa_{abc} \to (-1)^{n_x}\kappa_{abc}$ where $n_{x}$ counts the number of $x$-indices. Both our models break $M_{x}$ through the $\lambda \sin k_y$ term. Mirror symmetry $M_{y}$ is preserved. This forces the components $\kappa_{xxy}$, $\kappa_{yxx}$, and $\kappa_{yyy}$ to vanish. This restriction leaves only $\kappa_{xxx}$, $\kappa_{xyy}$, and $\kappa_{yxy}$ as permissible.

\subsection{Algebraic proof of $C_{2}$ symmetry protection}
Under a $C_{2}$ rotation ($\bk \to -\bk$), every spatial index acquires a minus sign, so that $\kappa_{abc} \to (-1)^{3}\kappa_{abc} = -\kappa_{abc}$. If the Hamiltonian preserves $C_{2}$, the macroscopic transport coefficient must obey $\kappa_{abc} \equiv 0$. The physical $C_{2}$ rotation acts as $\bk \to -\bk$ in momentum space. We use a pseudospin representation $U = \sigma_{z}$. This representation preserves the $\sigma_{z}$ altermagnetic splitting channel while flipping the $\sigma_{x}$ interband coupling channel. We test the unitary operator as follows:
\begin{align}
\sigma_{z} H(\bk) \sigma_{z} &= \sigma_{z} (g_{x}(\bk)\sigma_{x} + g_{z}(\bk)\sigma_{z}) \sigma_{z} \nonumber \\
&= g_{x}(\bk) (\sigma_{z} \sigma_{x} \sigma_{z}) + g_{z}(\bk) (\sigma_{z} \sigma_{z} \sigma_{z}).
\end{align}
We apply the Pauli matrix identities $\sigma_{z} \sigma_{x} \sigma_{z} = -\sigma_x$ and $\sigma_z \sigma_z \sigma_z = \sigma_z$. We find:
\begin{equation}
\sigma_{z} H(\bk) \sigma_{z} = -g_{x}(\bk)\sigma_{x} + g_{z}(\bk)\sigma_{z}.
\label{eq:matrix_proof}
\end{equation}
For the $g$-wave model, $g_{z}(-\bk)=g_{z}(\bk)$ and $g_{x}(-\bk)=-g_{x}(\bk)$. The Hamiltonian at the inverted momentum is $H(-\bk) = g_{x}(-\bk)\sigma_{x} + g_{z}(-\bk)\sigma_{z} = -g_{x}(\bk)\sigma_{x} + g_{z}(\bk)\sigma_{z}$. Comparing this result with Eq.~(\ref{eq:matrix_proof}), we find the exact equivalence $\sigma_{z} H(\bk) \sigma_{z} = H(-\bk)$. The $g$-wave system perfectly preserves $C_{2}$ symmetry. It mathematically forces $\kappa_{xyy} = 0$. In contrast, the $d$-wave model contains a parity-even term ($2t_{1}\sin k_{x}\sin k_{y}$) in $g_{x}(\bk)$. This means $g_{x}(-\bk) \neq \pm g_{x}(\bk)$, the unitary equivalence breaks down, $C_{2}$ is broken, and $\kappa_{xyy}$ survives.

\section{Formalism and Integration by Parts}
We employ Luttinger's formal prescription to relate the thermal transport coefficient to the electrical Berry connection polarizability (BCP) $\mathcal{G}_{xy}^{(n)} = n g_{xy} / (2G)$ \cite{Luttinger1964, Gao2014}. The substitution $e\mathbf{E} \to \frac{\epsilon-\mu}{T}\nabla T$ maps the electrical nonlinear Hall conductivity onto its thermal counterpart. This formal correspondence is rigorously established in the linear regime by Luttinger \cite{Luttinger1964} and Qin, Niu, and Shi \cite{Qin2011}. In the nonlinear regime, the analogous electrical BCP formula has been experimentally validated \cite{Gao2023}. A fully rigorous quantum kinetic theory incorporating magnetization current corrections \cite{Qin2011, Xiao2006} and second-order positional shifts remains an open challenge. We focus here on the leading-order geometric contributions which dominate the symmetry-protected transport bounds.

The raw integral takes the form:
\begin{equation}
\kappa_{xyy} = \frac{1}{\hbar T^{2}}\int_{\BZ}\frac{d^{2}k}{(2\pi)^{2}}\sum_{n} W(\epsilon_{n}) \partial_{k_{y}} \mathcal{G}_{xy}^{(n)},
\end{equation}
where $W(\epsilon) = (\epsilon-\mu)^{2} f_{0}(\epsilon)$ represents the thermal weight. Near the nodal lines where the energy gap approaches zero, the polarizability behaves as $\mathcal{G}_{xy} \propto 1/G^{3}$ and its derivative scales as $1/G^{4}$, as can be seen directly from Eq.~(\ref{eq:qm_derived}). We tame this mathematical singularity by performing integration by parts over the Brillouin zone:
\begin{equation}
\int \left( \partial_{k_{y}} \mathcal{G}_{xy} \right) W(\epsilon) d^{2}k = - \int \mathcal{G}_{xy} \left( \partial_{k_{y}} W(\epsilon) \right) d^{2}k.
\end{equation}
We apply the chain rule $\partial_{k_{y}} = \hbar v_{y} \partial_{\epsilon}$. We expand the energy derivative of the thermal weight using the product rule:
\begin{equation}
\frac{\partial W(\epsilon)}{\partial \epsilon} = 2(\epsilon-\mu) f_0( \epsilon) + (\epsilon-\mu)^{2} f_0'( \epsilon).
\end{equation}
Substituting this expansion back into the transport equation yields the robust and divergence-free formula:
\begin{equation}
\kappa_{xyy} = -\frac{1}{\hbar T^{2}}\int_{\BZ}\frac{d^{2}k}{(2\pi)^2}\sum_{n} \mathcal{G}_{xy}^{(n)} W_n'(\epsilon) v_{y,n}.
\label{eq:kappa_smooth}
\end{equation}
This step-by-step unrolling shifts the derivative onto the smooth Fermi-Dirac distribution.

\begin{figure}[t]
\centering
\includegraphics[width=\columnwidth]{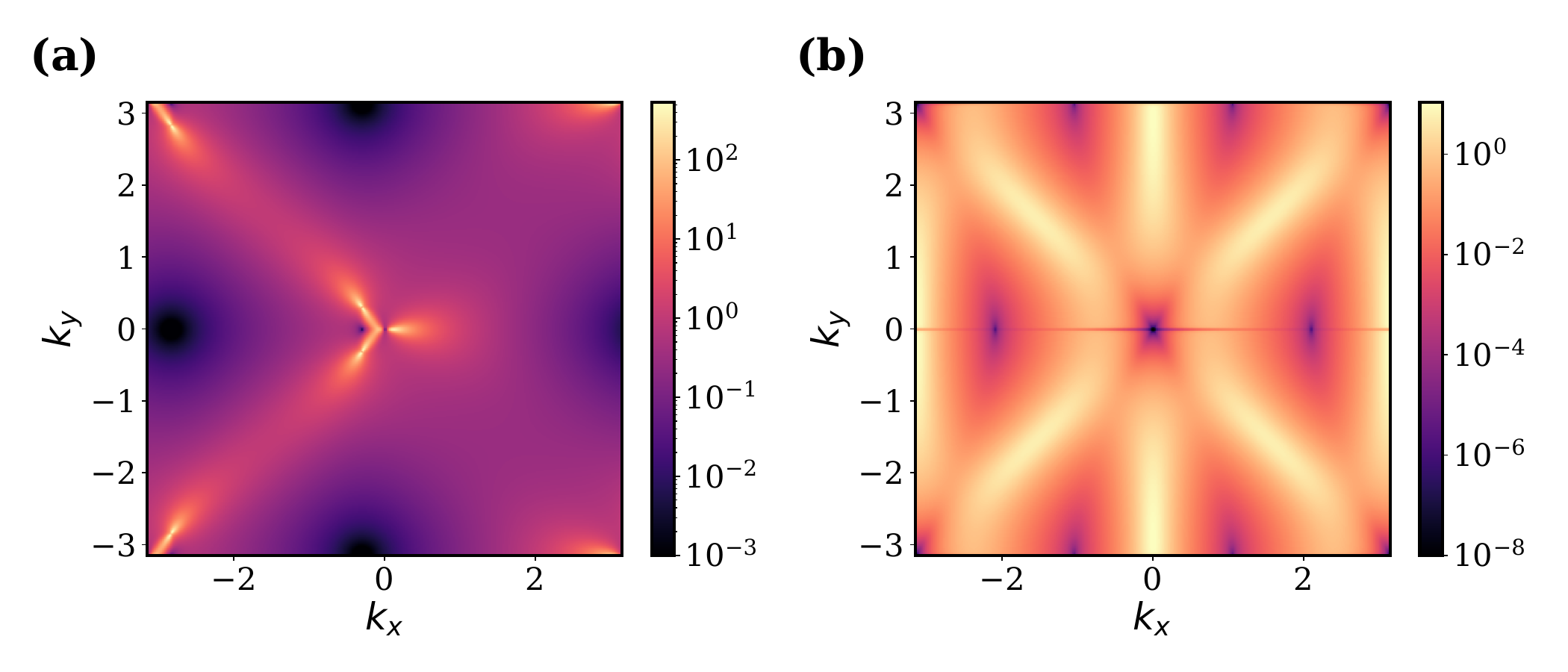}
\caption{Distribution of the trace of the quantum metric, $\mathrm{Tr}(g_{ab})$. (a) $d$-wave and (b) $g$-wave models form geometric hotspots near nodal lines. These singular structures, shown on a logarithmic scale, act as the source for macroscopic transport. Parameters: $\Delta = 1.0$, $t_{1} = 0.5$, $\lambda = 0.3$.}
\label{fig:metric_trace}
\end{figure}

\begin{figure}[t]
\centering
\includegraphics[width=\columnwidth]{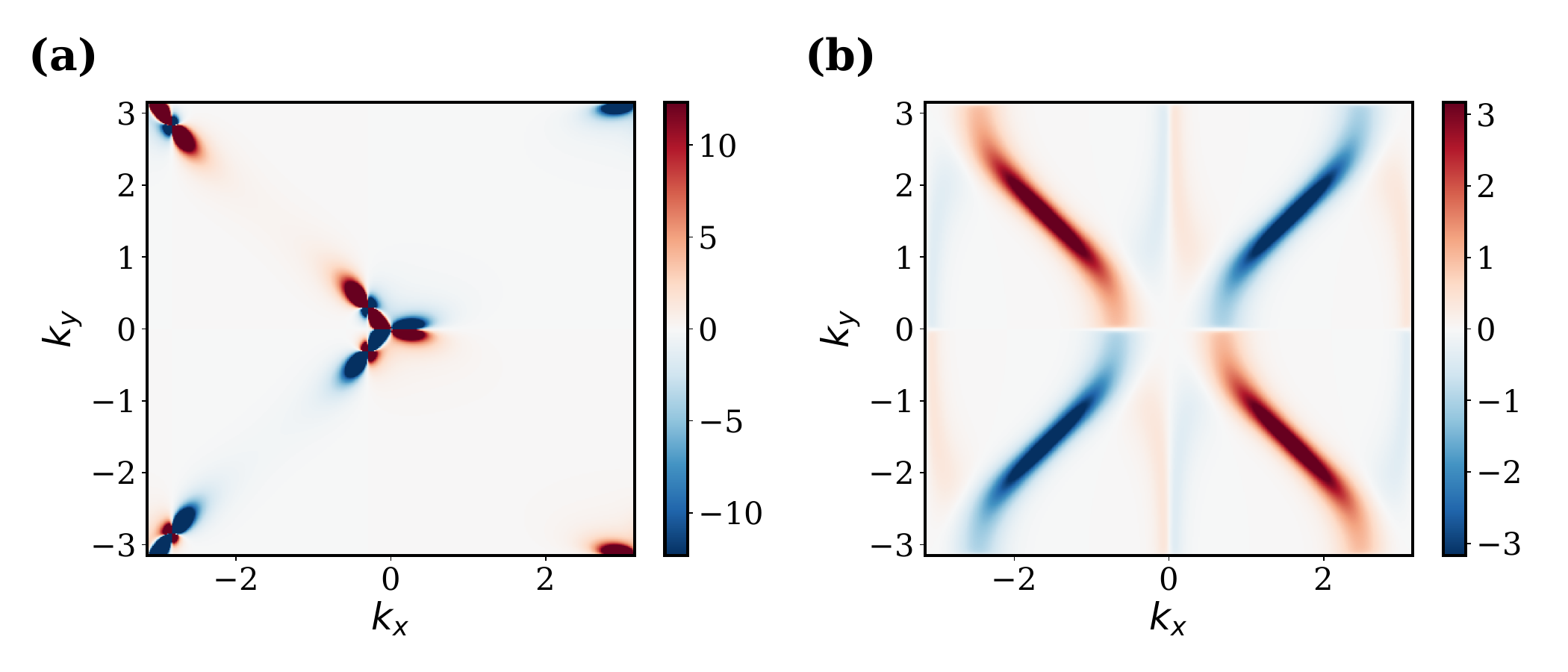}
\caption{Contrast in the symmetry of the Berry connection polarizability $\mathcal{G}_{xy}$. (a) The asymmetric texture of $d$-wave indicates $C_{2}$ breaking. (b) The perfectly antisymmetric structure of $g$-wave visually proves the protection by $C_{2}$ symmetry. This point-symmetric structure forces the global integral to vanish. Parameters: $\Delta = 1.0$, $t_{1} = 0.5$, $\lambda = 0.3$.}
\label{fig:bcp}
\end{figure}

\section{Analytical Scaling and Numerical Results}

\subsection{Analytical scaling relation}
We extract the parametric dependence of the conductivity for the $d$-wave model. We expand the product of derivatives $\varphi_{x} \varphi_{y}$ to first order in the symmetry-breaking parameter $\lambda$. This reveals the relation $\varphi_{x} \varphi_{y} \approx \mathcal{F}(\bk) \cdot (\lambda \Delta t_{1})$. Since the function $\mathcal{F}(\bk)$ is a parity-even function of momentum, it survives the Brillouin zone integration without cancellation. This results in the linear scaling law $\kappa_{xyy} \propto \lambda \Delta t_{1}$.

\subsection{Numerical verification}
We compute $\kappa_{xyy}$ using adaptive sub-grid patching to resolve nodal singularities. The parameters are $\Delta=1.0$, $t_{1}=0.5$, $\mu=-0.5$, $T=0.05$, and broadening $\eta=0.005$.

\begin{figure}[t]
\centering
\includegraphics[width=\columnwidth]{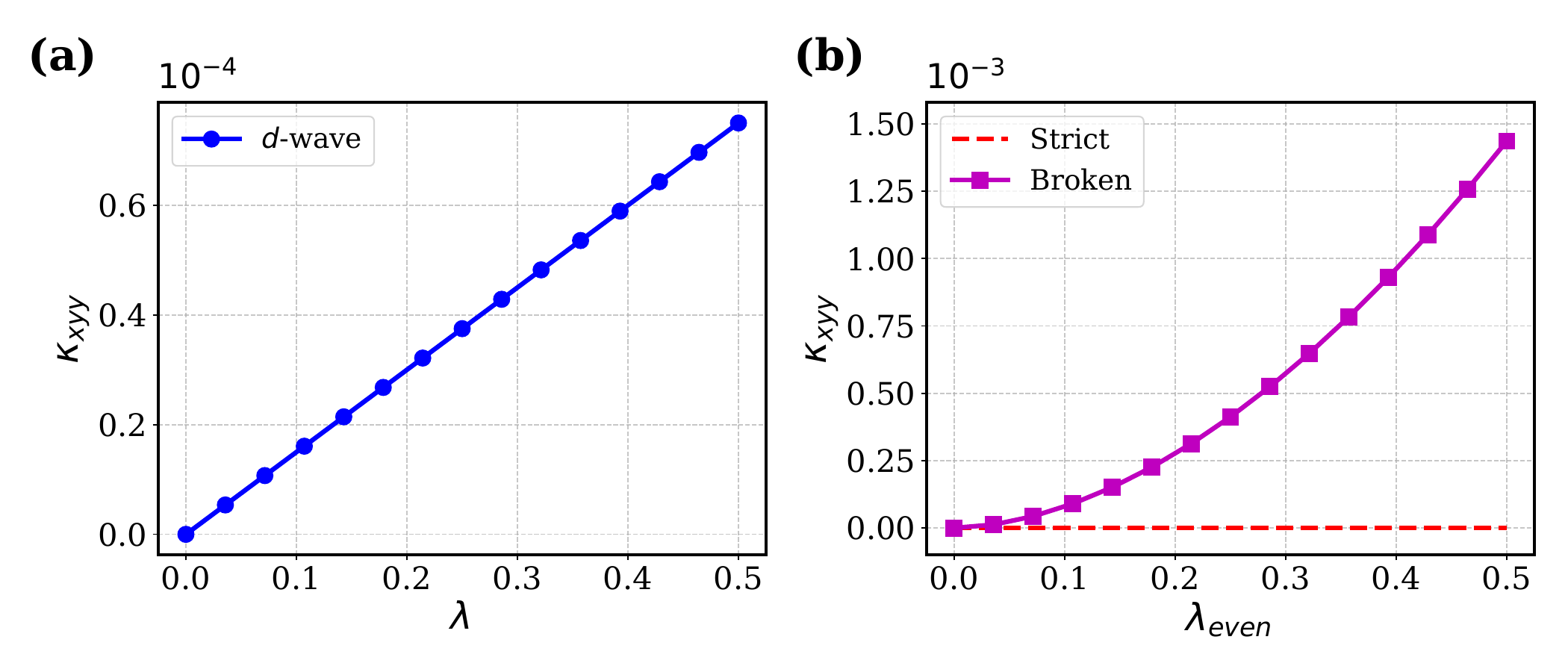}
\caption{Numerical evaluation of the nonlinear thermal Hall conductivity $\kappa_{xyy}$. (a) The $d$-wave model displays a linear increase with the parameter $\lambda$. (b) The $g$-wave model stays exactly at zero when $C_{2}$ is preserved (dashed line). The signal recovers sharply when $C_{2}$ is broken (solid line with squares). Parameters: $\Delta = 1.0$, $\mu = -0.5$, $T = 0.05$, $\eta = 0.005$; $t_{1} = 0.5$ for $d$-wave.}
\label{fig:kappa}
\end{figure}

For the $d$-wave model, $\kappa_{xyy}$ exhibits a clean, monotonic linear dependence on the mirror-breaking parameter $\lambda$ [Fig.~\ref{fig:kappa}(a)], confirming the analytically derived scaling.

The $g$-wave model results [Fig.~\ref{fig:kappa}(b)] serve as the numerical proof of our symmetry theorem. When $C_{2}$ is preserved ($\lambda_{\text{even}} = 0$), $\kappa_{xyy}$ remains locked at numerical zero ($\sim 10^{-19}$). Immediately upon introducing an even-parity, $C_{2}$-breaking term to the interband coupling ($\lambda_{\text{even}} \neq 0$), a robust geometric signal is recovered, increasing monotonically to match the scale of the $d$-wave response.

\section{Conclusion}
We have established the fundamental selection rules for the intrinsic nonlinear thermal Hall effect driven by the quantum metric in altermagnets. The twofold rotational symmetry ($C_{2}$) acts as the primary gatekeeper for the macroscopic geometric response. Through rigorous mathematical derivations and numerical verification, we demonstrated that $d$-wave materials naturally generate a finite signal by breaking this rotational protection, while $g$-wave systems preserve $C_{2}$ symmetry and extinguish the geometric signal.

We emphasize the generality of our central result. The $C_{2}$ selection rule is lattice-independent: it is a consequence of point-group symmetry acting on a rank-3 tensor, valid for any lattice geometry including the hexagonal structures of actual $g$-wave altermagnets such as CrSb and NiS. Whether a specific material preserves or breaks $C_{2}$ depends on the microscopic details of the spin-orbit coupling, which must be determined by first-principles calculations for each candidate.

These selection rules provide a diagnostic tool for current experimental controversies. If ideal RuO$_{2}$ with a pristine rutile structure preserves $C_{2}$ symmetry, the intrinsic nonlinear thermal Hall effect must be zero \cite{Smejkal2022b}. A finite observed signal would provide evidence of hidden symmetry breaking such as spin canting or defect-induced strain. Conversely, Mn$_{5}$Si$_{3}$ is an optimal platform among $d$-wave altermagnets for observing these geometric responses, because its orthorhombic distortion below the N\'{e}el temperature naturally breaks $C_{2}$ \cite{Reichlova2024, Han2025}. Our results clarify the microscopic origin of nonlinear thermal transport and provide guidelines for designing spin-caloritronic devices.

\end{document}